# Compact Graphene Plasmonic Slot Photodetector on Silicon-on-insulator with High Responsivity

*Zhizhen Ma[1], Kazuya Kikunage[1], Hao Wang[1], Shuai Sun[1], Rubab Amin[1], Mohammad Tahersima[1], Rishi Maiti[1], Mario Miscuglio[1], Hamed Dalir[2], Volker J. Sorger[1]*

1. Department of Electrical and Computer Engineering, George Washington University, 800 22nd St., Washington, District of Columbia 20052, USA
2. Omega Optics, Inc. 8500 Shoal Creek Blvd., Bldg. 4, Suite 200, Austin, Texas 78757, USA

**ABSTRACT**

Graphene has extraordinary electro-optic properties and is therefore a promising candidate for monolithic photonic devices such as photodetectors. However, the integration of this atom-thin layer material with bulky photonic components usually results in a weak light-graphene interaction leading to large device lengths limiting electro-optic performance. In contrast, here we demonstrate a plasmonic slot graphene photodetector on silicon-on-insulator platform with high-responsivity given the 5 µm-short device length. We observe that the maximum photocurrent, and hence the highest responsivity, scales inversely with the slot gap width. Using a dual-lithography step, we realize 15 nm narrow slots that show a 15-times higher responsivity per unit device-length compared to photonic graphene photodetectors. Furthermore, we reveal that the back-gated electrostatics is overshadowed by channel-doping contributions induced by the contacts of this ultra-short channel **graphene photodetector. This leads to quasi charge neutrality, which explains both the previously-unseen offset between the maximum photovoltaic-based photocurrent relative to graphene's Dirac point and the observed non-ambipolar transport. Such micrometer compact and absorption-efficient photodetectors allow for short-carrier pathways in next-generation photonic components, while being an ideal testbed to study short-channel carrier physics in graphene optoelectronics.**

**Introduction.**

Graphene has become a complementary platform for electronics and optoelectronics because of its remarkable properties and versatility(*1*). A variety of applications exploit graphene's peculiar features to include modulators(*2*), plasmonic optoelectronics(*3–6*), photovoltaic devices(*7*), ultrafast lasers(*8*), and photo-detection(*9, 10*). For photo conversion applications the linear and gap-less band structure of graphene results in wavelength-independent absorption(*11, 12*). Moreover, graphene's carrier can be tuned via electrostatically doping, thus modulating light absorption. Due to its superb carrier mobility(*13, 14*), graphene-based absorption enables ultrafast conversion of photons or plasmons to electrical currents or voltages. However, the light-graphene interaction, and consequently the responsivity of graphene-based devices, is usually rather weak due to the geometrical mismatch between graphene's atom-thin thickness and the diffraction-limited optical mode area of photonic components.

The first-generation of graphene-based free-space photodetectors (PDs) uses metal-graphene-metal structures(*14*); choosing different work-functions for the source- and drain contacts results in an asymmetric band structure, thus enabling non-biased band-bending for charge polarity separation, leading to near-zero dark current. Interdigitated metallic contacts, are typically adopted to reduce the required photodetector area(*14*, *15*). Integrating graphene with with colloidal QDs(*16*) and plasmonic structures(*17*) enhances the PD's responsivity due to increased local density of states and increased optical path-length induced by particle scattering.

Nevertheless, the most used PD schemes in photonic integrated circuits to date are still based on Ge and InP(*18*, *19*), which are, however, intrinsically challenged by poor electro-optic tunability and reduced carrier dynamics often induced by low surface velocity recombination from the etch process. In this context graphene has emerged as an alternative active-material for optoelectronic components including PDs, due to its superior electro-optics properties, i.e. tunable optical properties, fast carrier dynamics, broadband functionalities, and ease-of-integration enabled by wafer-scale growth processes. To increase the absorption in graphene and mode overlap integrated metal graphene−silicon PDs have been investigated(*2*, *10*, *20*, *21*). Here the temporal response time (3dB roll-off speed) is limited by the carrier transit time due typically long channel lengths(*22*). In this work we argue and show that the performance of graphene-metal based PDs can be optimized by enhancing the light-graphene interaction through exploiting the unique carrier characteristics inside plasmonic short-channel transport devices.

Here, we report the design and characterization of a plasmonic slot graphene photodetector monolithically integrated on silicon-on-insulator. While our plasmonic slot design results in a comparable (absolute) responsivity compared to that of plasmonic graphene detectors recently reported by Ding et al.(*23*), the underlying physics and relative performances are substantially different as discussed below; in brief our design utilizes hybrid slot plasmonic mode with a smallest gap-size (15 nm) to date. Interestingly, the maximum photo-absorption is achieved for smallest plasmonic gap. To understand the short-channel effect of this novel design, we investigate a symmetric metal-contact work-function concept, which therefore requires a bias voltage to extract the photo carriers. However, the generality of our observed results and short-channel explanation hold true for both the symmetric and the asymmetric metal work-function case. Our structure sits on top of the silicon-on-insulator (SOI) epi layer. Besides providing a different integration method to the plasmonic graphene PD with conventional photonic integrated circuit, the silicon layer can be used as a back gate to achieve two device operation regimes (i.e. bolometric effect or photovoltaic effect). Also, this PD design allows for dual operation (positive or negative photocurrent) depending on the gate voltage. Multiple devices with a variety of geometric dimension have been fabricated and studied, and the device featuring the smallest plasmonic slot gap size (15 nm) showed the highest responsivity of 0.35 A/W while being only 5 μm short and a bias of 0.2 V. This is about 15-times more efficient per device length than integrating graphene onto an SOI waveguide(*22*), enabled by the strong light-graphene interaction enhancement from the narrow plasmonic slot structure. As expected this device also shows broadband (here 100 nm tested) operation. We believe this compact yet high-responsivity photo-detector enables next-generation optoelectronics specifically for both dense integration and short temporal delays.

**Results**

Graphene is has a strong anisotropic material permittivity tensor that shows a high in-plane electric field absorption with respect to its lattice plane while the out-of-plane permittivity is similar to graphite, thus not significantly contributing to light absorption(*24–26*). It is therefore of interest for graphene electro-optic devices to enhance the in-plane field confinement, which we achieve be selecting a metal (plasmonic) slot-based design. This PD uses patterned CVD-grown graphene with the plasmonic slot placed on top of the SOI substrate bridged by a 15 nm $SiO_2$ layer serving as the spacer to a) support the optical mode (hybrid gap plasmon), and b) to tune graphene's Fermi-level (Fig. 1 (a) and (b)). Note that the graphene film actually extends outwards from the slot region for a few hundred of nanometers (Fig. 1 (c) and (d)) to ensure a uniform substrate profile for the dose-sensitive slot region during the electron-beam lithography (EBL) process. The metallic grating coupler, taper and slot waveguide are made of Ti/Au of 2 nm/48 nm in a two-steps process (see method section). A relatively short (50 nm) slot waveguide height was chosen to allow for sufficient light confinement. For device test, linearly TE-polarized light is coupled into the chip through the metallic grating coupler which then adiabatically funnels the mode tapered region to the plasmonic slot waveguide section(*27*). When the slot waveguide gap size is small enough (below 50 nm), light coupled inside the plasmonic slot region propagates predominantly as hybrid plasmonic mode and evanescently travels along the slot waveguide with a minimized mode area(*28*, *29*). We selected this mode since it provides a strong in-plane field confinement inside the slot region enabling enhancement of graphene's absorption. In fact we find the photoresponse improving with a reduction in gap width ($W$), which will be discussed in next section (Fig. 1(b)). Foreshadowing the detailed discussion below, this ultra-narrow ($W = 15$ nm) slot waveguide provides a thermal hot-spot inside the gap due to the enhanced light absorption from both graphene and metallic sidewall, thus a substantial temperature-dependent bolometric effect is anticipated and indeed observed (Fig. 3c). Lastly, the small gap size provides a short drift path for photon-generated carriers to be collected by the electrodes when the device is operated in the photovoltaic effect regime, which also reduces the chance of electron-hole recombination(*23*, *24*).

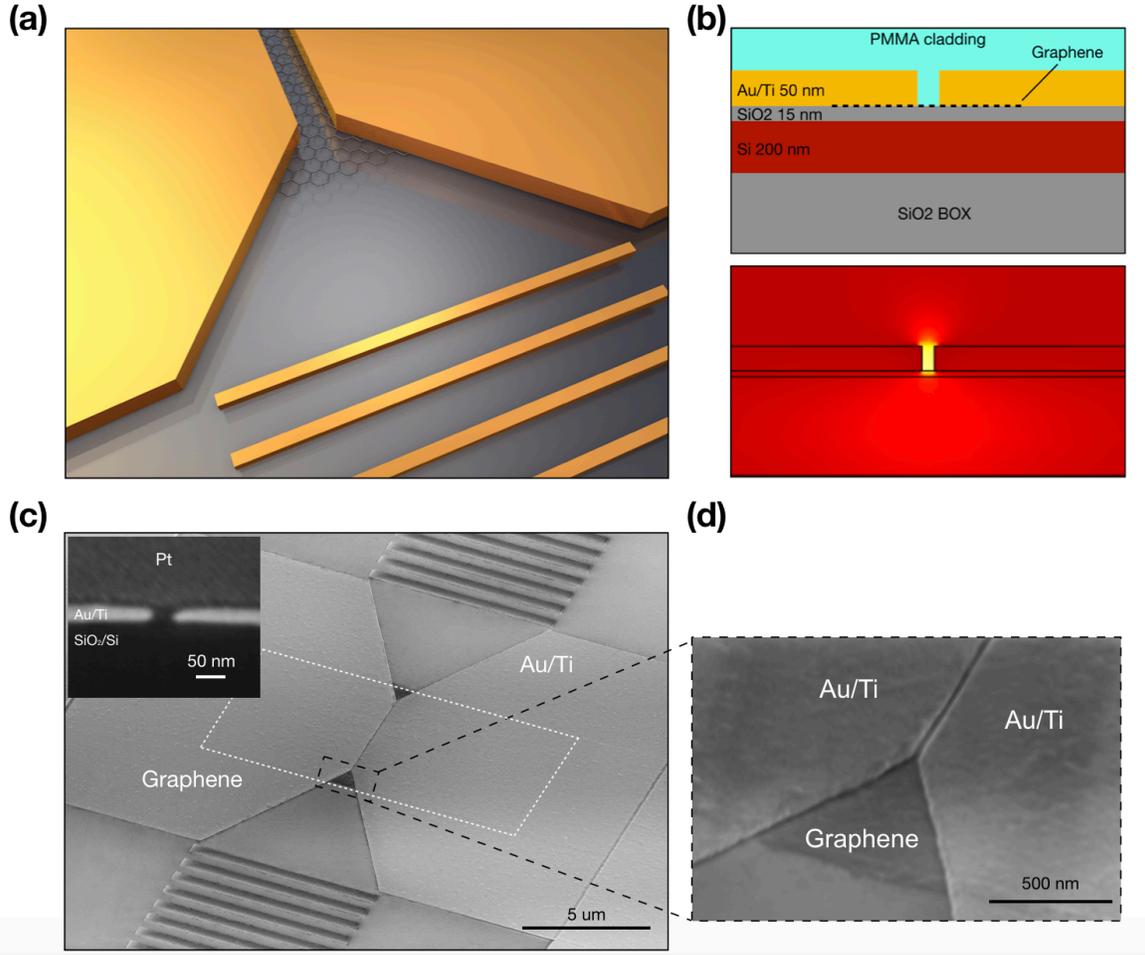

Figure 1. (a) 3-dimensional representation of the photodetector. (b) Cross sectional schematic of the device, where the Ti/Au metallic structures are in close proximity to each other for forming the plasmonic slot waveguide, as well as serving as the source-drain contact, while the p-doped SOI device layer silicon is used for back gate with a thin silica layer in between. (c) SEM image of fabricated photodetector, graphene is underneath the metal structure and the cross sectional image as inset, here is a slot with slot gap width $W = 30$ nm and length $L = 5\mu$m. (d) Zoom in view of the tapered region and slot, note that the graphene layer is slightly extended out from the slot region for a few hundred of nanometer to ensure good alignment.

We perform Raman spectroscopy on fabricated devices to probe the quality of graphene underneath the slot structure for possible impact during fabrication. The Raman spectrum shows the expected monolayer graphene feature where both G and 2D peaks are preserved even with the slot size down to ~15 nm, though a reduced signal-to-noise ratio (SNR) is observed with reduction in gap size (Fig. 2 (a)). We attribute both the small 2D peak shift (less than ten cm$^{-1}$) and the reduced $I_{2D}/I_G$ ratio (approaching unity) for decreasing gap size to the direct deposition and lift-off of metal on top of graphene(*30*), which induces random surface defects and PMMA residues on graphene sheet that cause minor degradation and mild p-type doping(*31–33*). The Raman laser profile along with the device structure attests that the pump laser (532 nm) power is primarily focused on the slot region thus confirming the measured Raman signal has to be attributed to the graphene inside the gap (Fig. 2 (b)). Incidentally, our plasmonic slot waveguide introduces a localized surface plasmon (LSP) resonance which further enhances the pump beam where graphene resides, thus contributing to an enhanced Raman signal(*34*), which allows to resolve even the narrowest (15 nm) channel graphene layer within the slot.

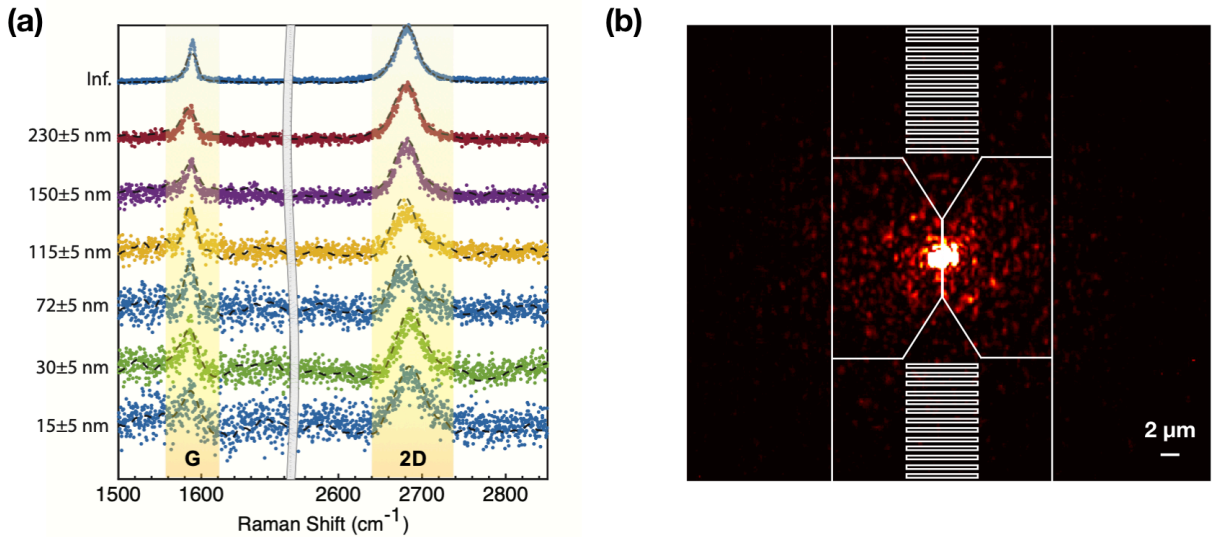

Figure 2. (a) Raman shift for samples with different slot gap size. Strong 2D and G peak indicates that the Graphene is preserved after the fabrication process. A change of doping level with varying gap width is observed, which we believe is mainly due to the increased defect and PMMA residue with smaller gap size. (b) Raman laser profile and the outline of device under measurement, note that the laser spot size is smaller than the gap length, which indicates that the Raman signal measured came from graphene inside the plasmonic slot.

For graphene-based photodetectors, three different compelling mechanisms contribute to the photo-response, namely photo-thermolelectric (PTE) effect, photo-bolometric (PB) effect and photovoltaic (PV) effect. The PTE effect relies on the photon induced electron temperature difference between two different graphene doping regions, which is usually achieved by designing an asymmetric band diagram from different material doping or spatially placing the photosensitive region away from the symmetric part (*35*). However, in our design, the same metal combination (Ti/Au) is used for the plasmonic slot which is simultaneously used to contact graphene. This results in a symmetric band diagram across the active region, and therefore minimizes the contributions from the PTE (Fig. 3). Instead, a competing PB and PV effect is observed due to their inverse photocurrent polarity; that is, the PB contributes to a negative photocurrent due to the increased channel resistance from smaller mean free path for heated carriers, whereas photo-generated carriers from PV reduce the channel resistance. By varying the gate voltage to change the channel doping level, two distinct PB dominant vs. PV dominant regions were found for our devices (Fig. 3). Furthermore, this short-channel detector shows that the PV generated photocurrent peak (minimized PB) does not coincide with the same gate bias where the Dirac point. This is due to the metal-induced short-channel doping of the graphene sheet, which is no longer a negligible effect for devices with such small channel length, i.e. the channel length (width *W* of the plasmon gap) is shorter than the charge transfer region $L_{ct}$ (Fig. 3a).

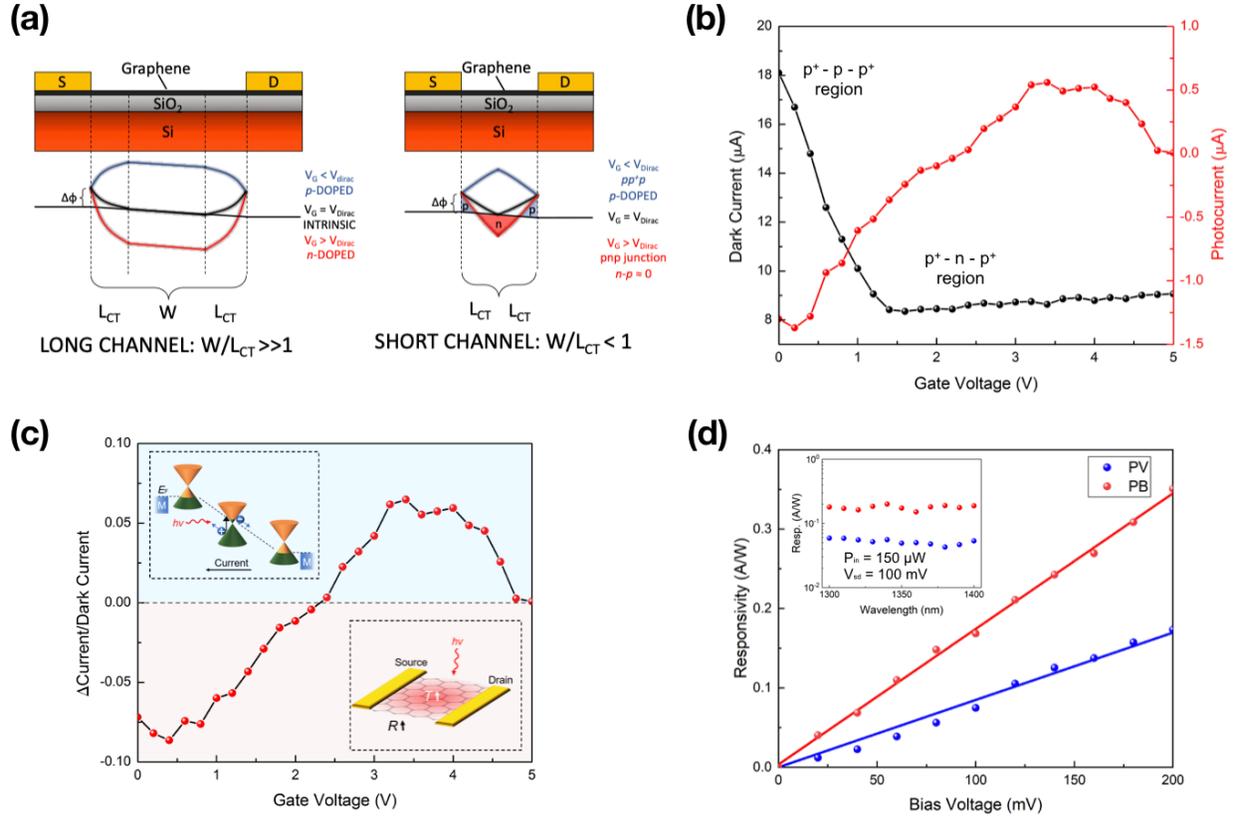

Figure 3. Device performance. (a) Band diagram for long and short channel device. For the short-channel detector, because of the small $W/L_{ct}$, the charge transfer region has a significant influence inside the channel. (b) Measurement of photocurrent and dark current ($I_{sd}$) from the photodetector ($V_{sd} = 0.1V$), the photocurrent changes sign when the photovoltaic effect is stronger than bolometric effect (c) Change of current, two peaks are observed in different polarity for either bolometric effect and PV effect dominant regions, both with a maximum ratio of more than 7% (d) Measured responsivity vs. bias voltage, a linear fit of 1.71A/WV and 0.85A/WV for PB and PV effect is retrieved, respectively. Inset shows a broadband responsivity from 1300 to 1400 nm. The measurement is mainly limited by metallic grating coupler operating wavelength.

Unlike previous graphene PDs using plasmonic structures, here we report results based on a short-channel devices, highlighted exemplary in Figure 3 (i.e. $W$ = 15 nm, $L$ = 5 µm). Sweeping the gap width (15 - 250 nm) we find the highest photo-response for the smallest gap width despite its narrowest absorption region (least amount of graphene), which is a key design feature of the plasmonically-enhanced field confinement. During measurement, a 8 µW optical signal at 1310 nm wavelength was coupled into the plasmonic slot structure, and the source-drain current and photocurrent under different gating voltage was measured with a static bias voltage of 0.1 V (Fig. 3b). Firstly, we notice that the source-drain current is different from a typical graphene's ambipolar

I-V curve relative to the Dirac point (~1.4 V), which is a sign of a strong asymmetric charge transfer in this short-channel device. It is known that the graphene sheet underneath the metal contact has a gate-uncontrollable charge density due to Fermi level pinning, which defines the hundreds of nanometers-wide charge transfer region into the channel(*36*, *37*). Thus, for long channel ($W >> L_{ct}$) graphene transistor-like devices, the Fermi level pinning usually introduces a minor degraded carrier transport performance when the channel is electrostatically gated from one to the other polarity, due to the formation of *p-n* junctions inside the channel (Fig. 3(a) shows a p-type pinning and red-dashed line indicates n-doped channel)(*38*, *39*). However, for this short-channel PD the channel length is actually smaller than the charge transfer region ($W << L_{ct}$). Thus, when the gating voltage is larger than $V_{Dirac}$ (n-type doping inside channel), an increased ballistic transport contribution gives rise to a large excessive resistance which stems from the selective transmission of carriers through the *p-n-p* junction, with only carriers approaching the barrier in an almost perpendicular direction allowed for passage(*37*, *40*). Also, our 3-termincal transport data on devices with different gap size shows that a larger channel size (wider slot gaps) are less affected by this excessive resistance, which indicates a transition from ballistic regime to diffusive transport regime (see supplementary information).

Without applying back gate voltage, the measured photocurrent shows a strong bolometric effect, which can be attributed to heating up of the carriers inside the channel and thus inducing excessive resistance due to reduced mean free path of hot carriers (Fig. 3(b), red line). Upon back-gating the PB effect decreases with increasing $V_g$, due to reduced carriers inside the channel, until the PV effect peaks around $V_g = 3.4$ V, with a corresponding photocurrent of 0.52 µA. It is worth noticing that the PV generated photocurrent peak, unlike reported in Ref.(*41*), is not in the vicinity of the Dirac point. This can also be explained by the short-channel device model where the channel is *p*- *i-p* doped when the Dirac point gating voltage is reached, due to the existence of charge transfer region. Here, the holes in the charge transfer region are no longer negligible because of the short-channel size and the limited density of states (DOS) in graphene, thus the net carrier inside this photo-sensitive channel is still away from neutrality. Only when $V_g > V_{Dirac}$, the graphene sheet is *n*-doped to a level that the band edge of graphene crosses the Fermi level twice as the formation of a *p-n-p* junction, the net carrier inside the channel could approach zero thus resulting a minimized PB effect. After this point, the photocurrent reduces again since increasing electrons become the majority carriers inside the channel, and are subject to the continuous increase of $V_g$, eventually the photocurrent crosses zero again. These results are in general good agreement with the six-folded graphene photoresponse pattern as reported in previous studies(*2, 9, 41*).

As reported(*9*), it is usually recommended to not use bias voltage for graphene based photodetector, since this leads to a higher dark current, thus lowering the SNR. However, for our design, the active region has an almost symmetric band structure, thus near zero responsivity was observed for unbiased cases. With a limited bias of 0.1 V, the ratio between change of current and the dark current is explicitly plotted in Fig. 3c; here two distinct regions based on the polarity of photocurrent are observed, where the negative (positive) change of current is due to PB (PV) effect. A substantial SNR of 9% was achieved for PB effect peak around zero gating while 7% of SNR was observed for the peak PV effect. We notice that the excessive resistance for the *p-n-p* doped channel actually helps suppressing the dark current, thus boosting the signal-to-noise ratio for PV effect dominant region.

The bias-dependent responsivity for both PB effect and PV effect, under proper gating, shows that the external responsivity for PB generated photocurrent could reach 0.35 A/W, with 0.17 A/W for PV generated photocurrent, under a small bias of only 0.2 V (Fig. 3d). Here a quasi-linear relation can be extrapolated for bias voltage and responsivity, indicating that a higher responsivity can be achieved by still increasing the bias. However, since the dark current also increases linearly with applied bias, SNR does not increase with bias. Instead, we focus on the slope of this dependency as a steeper slope leads to a more sensitive performance with respect to the bias. For our device, a slope of 1.71 A/WV and 0.85A/WV was fitted for PB and PV effect, respectively, which is higher than the state-of-art bolometric or photovoltaic effect driven plasmonic graphene photodetector reported in Ref. (*23*, *24*). We attribute this efficient bias to our extremely small slot gap width, where both PB and PV effect are enhanced because of both high absorption and short separation path for carriers. Also, the expected broadband response is verified across 100 nm spectrum and shows a flat response in responsivity (inset, Fig. 3 (d)). Note that the measurement here was mainly limited by the metallic grating coupler spectral bandwidth, and an even broader response is expected given graphene's linear and bandgap-less band structure.

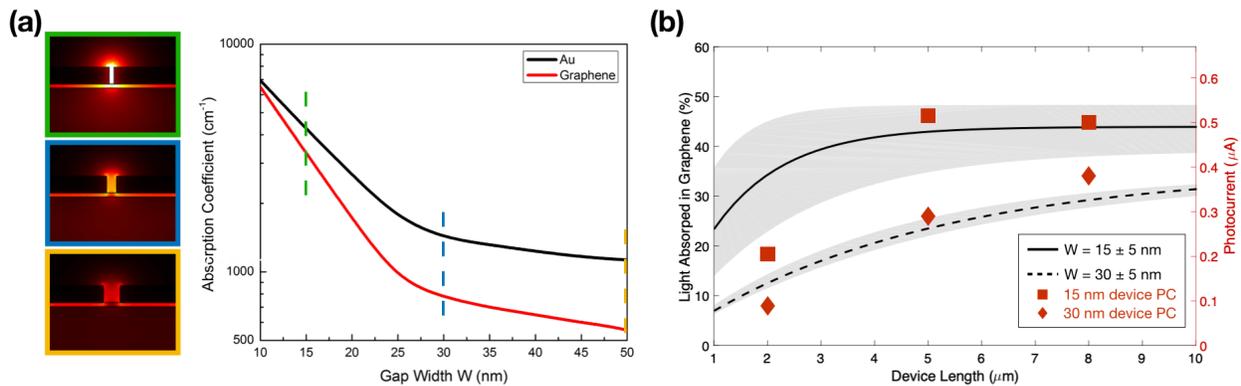

Figure 4 (a) Absorption coefficient of metal and graphene in the reported plasmonic slot detector with varying gap width. The absorption in graphene decreases quickly with increased gap size, due to mode leakage into silicon layer underneath the dielectric. Inset shows the E-field distribution for a plasmonic slot with $W$ = 15 nm, 30 nm and 50 nm, respectively. (b) Light absorption in graphene for different device length and the measured photocurrent.

It is well understood that plasmonic structures allow reducing the mode area (and volume), thus enhance the light-graphene interaction, however, with smaller gap size, a higher ohmic loss from the surface plasmon polariton was reported with reduced plasmonic waveguide propagation length(*27*), and is not ideal for such light-harvesting device. So, here we discuss the absorption relation between graphene and metal in our structure, and correlate the simulation result to our experimental measurements. For our plasmonic slot structure, devices with multiple *L* (2, 5, and 8 µm) and *W* (~15 - 250 nm) were fabricated and studied. However, for devices with slot size larger than 50 nm, no significant photocurrent was measured (see supplementary information). This is because when the plasmonic slot size is larger than 50 nm, the TE-like hybrid plasmonic mode, which promotes graphene absorption, is no longer the dominant mode due to the competing TM-like plasmonic mode having exceeded effective index, which causes enhanced TM-like mode inside the dielectric spacer and does not enhance graphene absorption, resulting in negligible photocurrent(*42*). Due to the weaker field confinement inside the gap with increasing gap size (transition from more plasmonic to more photonic for this hybrid mode), both absorption coefficient decays exponentially, however, the absorption in graphene decays more quickly compares to metal, which indicates that only when the gap size is small enough, the absorption of graphene could compare with metal (Fig. 4 (a)). With the absorption coefficient for graphene $\alpha_G$ and metal $\alpha_M$, the length dependency of fraction of light absorption in graphene $\eta$ can be calculated by(*43*)

$$\eta(L) = \frac{\alpha_G}{\alpha_M + \alpha_G} (1 - e^{-\alpha_G L} e^{-\alpha_M L}) \quad (1)$$

Using (1) to estimate the absorption in graphene for devices with 15 nm and 30 nm reveals that for a slot size of 15 ± 5 nm about 40% of light can be absorbed by graphene within a device length of only ~ 4 µm (Fig. 4 (b)). However, devices with a larger gap size of 30 ± 5 nm require more than 10 µm of device length to saturate the absorption in graphene (max ~30%), thus highlighting the importance of a small slot size for higher responsivity. Note, a gap size variation of ± 5 nm is assumed due to the deposited metal grain size and plotted as the shaded region. Our measured data points for photocurrent do agree well with the predicted gap-width and device-length trends (Fig. 4 (b)).

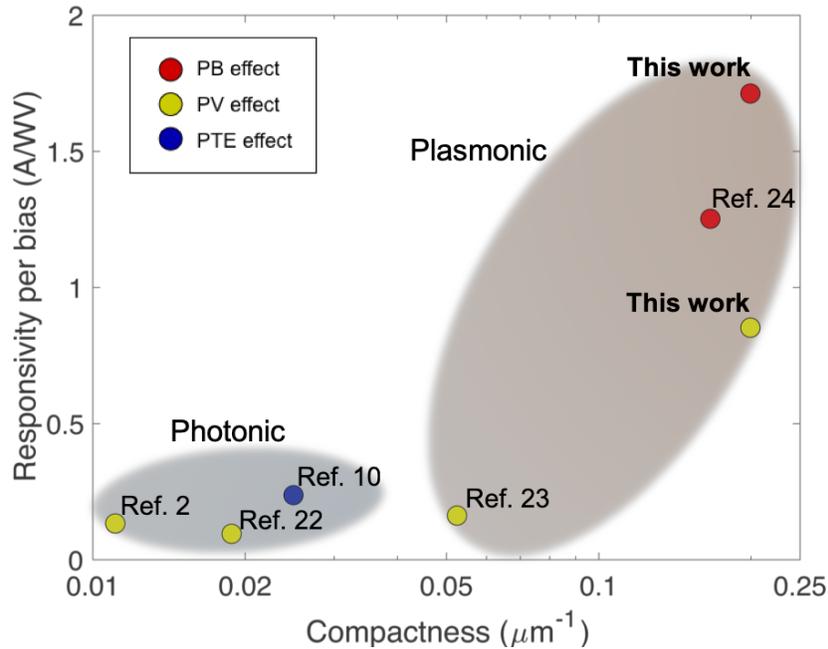

Figure 5 Device compactness (1/device length) and responsivity per bias voltage comparison of the short channel graphene photodetector with state-of-art graphene based photonic and plasmonic photodetectors, note that two distinct data point for our work represent either bolometric or photovoltaic effect generated photoresponse.

**Discussion.** Fig. 5 summarizes the state-of-art integrated graphene photodetectors, with the device compactness (1/device length) and responsivity per bias plotted. In general, the photonic graphene integrated PD requires a larger device footprint to achieve significant absorption in graphene, due to the dimension mismatch between graphene and dielectric waveguide hence a small mode overlap. In the plasmonic regime, for both PB and PV effect driven device, this ultra-short channel device demonstrated a higher responsivity per bias compare to recently reported graphene plasmonic detector(*23*, *24*), despite that our device has the smallest device length. We believe this superior performance in device compactness and responsivity per bias is because that our device features the smallest slot size *W* amongst these plasmonically enhanced PDs, thus a higher absorption in graphene and a smaller carrier drift path could be expected. In addition to the device demonstration, our photocurrent measurements provided considerable insightful details on the transport in short channel devices, highlighting the role of the metal-graphene junctions and the governing mechanism in the charge transport region. We measured that under different back-gate voltage, the ultra-short graphene channel, in fact, displayed a large mismatch between channel net charge neutrality point and the Dirac point for graphene sheet due to a significant influence of the metal contact doping into the ultra-short channel detector. This translated to a pronounced shift between the peak in PC and the Dirac Point. Moreover, the device performance, as function of the gap size of the slot, was investigated. Smaller gap size yielded to a stronger field confinement, which ultimately leads to a larger photoresponse.

In conclusion, hereby, we demonstrated an ultra-compact graphene based plasmonic slot detector with a dual responsivity of 0.35 A/W and 0.17 A/W for bolometric and photovoltaic effect, respectively. Moreover, our work provided an alternative approach for integrated graphene plasmonic optoelectronic device, and in the view of the detailed analysis of the transport mechanism in short channel, which paves the way for the engineering of ultra-short channel graphene plasmonic optoelectronic devices for integrated photonic circuits.

**Methods.** *Graphene Plasmonic Slot PD Fabrication.* The fabrication process for this device is depicted in Supporting info section 1. The device was fabricated on a commercially available silicon-on-insulator substrate (SOITEC) with a 200 nm p-type Si device layer on top of a 1000 nm buried oxide (BOX). By using atomic layer deposition (ALD), a 15 nm $SiO_2$ thin film layer was first deposited on top of the substrate as the gating oxide as well as the spacer layer to support the hybrid plasmonic slot mode(*27*). Next, monolayer graphene film (Graphenea, Easy Transfer) was wet transferred onto the substrate and then patterned by EBL (Raith Voyager) with negative photoresist (AR-N 7520) followed by oxygen plasma etching. After graphene patterning, for fabricating plasmonic slot with lower than 50 nm gap width, a first EBL step defined the grating coupler and left part for the metallic structure, followed by Ti/Au deposition and lift-off process. Then a second EBL process was performed to define the right part for the metal slot, followed by a second deposition and lift off process. This two-step process eases the lift-off process for small gap size (below 50 nm). A varying spatial separation was defined between the left and right metal structure to ensure having a variety of plasmonic slot size. Eventually, the Ti/Au contact pad was fabricated by using another EBL step followed by deposition and lift-off.

**Corresponding Author**

Volker Sorger, sorger@gwu.edu

**Funding Sources**

VS is funded by AFOSR (FA9550-17-1-0377) and ARO (W911NF-16-2-0194).

REFERENCES

1. F. Bonaccorso, Z. Sun, T. Hasan, A. C. Ferrari, Graphene photonics and optoelectronics. *Nat. Photonics*. **4**, 611 (2010).

2. N. Youngblood, Y. Anugrah, R. Ma, S. J. Koester, M. Li, Multifunctional Graphene Optical Modulator and Photodetector Integrated on Silicon Waveguides. *Nano Lett*. **14**, 2741–2746 (2014).

3. M. Miscuglio, D. Spirito, R. P. Zaccaria, R. Krahne, Shape Approaches for Enhancing Plasmon Propagation in Graphene. *ACS Photonics*. **3**, 2170–2175 (2016).

4. V. W. Brar, M. S. Jang, M. Sherrott, J. J. Lopez, H. A. Atwater, Highly Confined Tunable Mid-Infrared Plasmonics in Graphene Nanoresonators. *Nano Lett*. **13**, 2541–2547 (2013).

5. J. D. Cox, F. Javier García de Abajo, Electrically tunable nonlinear plasmonics in graphene nanoislands. *Nat Commun*. **5** (2014) (available at http://dx.doi.org/10.1038/ncomms6725).

6. F. J. García de Abajo, Graphene Plasmonics: Challenges and Opportunities. *ACS Photonics*. **1**, 135–152 (2014).

7. Plasmonics for improved photovoltaic devices | Nature Materials, (available at https://www.nature.com/articles/nmat2629).

8. Z. Sun *et al.*, Graphene Mode-Locked Ultrafast Laser. *ACS Nano*. **4**, 803–810 (2010).

9. F. H. L. Koppens *et al.*, Photodetectors based on graphene, other two-dimensional materials and hybrid systems. *Nat. Nanotechnol*. **9**, 780–793 (2014).

10. R.-J. Shiue *et al.*, High-Responsivity Graphene–Boron Nitride Photodetector and Autocorrelator in a Silicon Photonic Integrated Circuit. *Nano Lett*. **15**, 7288–7293 (2015).

11. A. K. Geim, Graphene: Status and Prospects. *Science*. **324**, 1530–1534 (2009).

12. A. K. Geim, K. S. Novoselov, The rise of graphene. *Nat. Mater*. **6**, 183 (2007).

13. F. Xia, V. Perebeinos, Y. Lin, Y. Wu, P. Avouris, The origins and limits of metal–graphene junction resistance. *Nat. Nanotechnol*. **6**, 179–184 (2011).

14. T. Mueller, F. Xia, P. Avouris, Graphene photodetectors for high-speed optical communications. *Nat. Photonics*. **4**, 297–301 (2010).

15. S. Cakmakyapan, P. K. Lu, A. Navabi, M. Jarrahi, Gold-patched graphene nano-stripes for high-responsivity and ultrafast photodetection from the visible to infrared regime. *Light Sci. Appl*. **7** (2018), doi:10.1038/s41377-018-0020-2.

16. G. Konstantatos *et al.*, Hybrid graphene–quantum dot phototransistors with ultrahigh gain. *Nat. Nanotechnol*. **7**, 363–368 (2012).

17. T. J. Echtermeyer *et al.*, Strong plasmonic enhancement of photovoltage in graphene. *Nat. Commun*. **2** (2011), doi:10.1038/ncomms1464.

18. J. Michel, J. Liu, L. C. Kimerling, High-performance Ge-on-Si photodetectors. *Nat. Photonics*. **4**, 527–534 (2010).

19. A. Beling, J. C. Campbell, InP-Based High-Speed Photodetectors. *J. Light. Technol*. **27**, 343–355 (2009).

20. G. Isić, R. Gajić, Lifetime and propagation length of light in nanoscopic metallic slots. *JOSA B*. **31**, 393–399 (2014).

21. J. Wang *et al.*, High-responsivity graphene-on-silicon slot waveguide photodetectors. *Nanoscale*. **8**, 13206–13211 (2016).

22. X. Gan *et al*., Chip-integrated ultrafast graphene photodetector with high responsivity. *Nat. Photonics*. **7**, 883–887 (2013).

23. Y. Ding *et al*., Ultra-compact graphene plasmonic photodetector with the bandwidth over 110GHz. *ArXiv180804815 Phys*. (2018) (available at http://arxiv.org/abs/1808.04815).

24. P. Ma *et al*., Plasmonically enhanced graphene photodetector featuring 100 GBd, high-responsivity and compact size, 11.

25. Z. Ma, M. H. Tahersima, S. Khan, V. J. Sorger, Two-Dimensional Material-Based Mode Confinement Engineering in Electro-Optic Modulators. *IEEE J. Sel. Top. Quantum Electron*. **23**, 81–88 (2017).

26. D. Ansell *et al*., Hybrid graphene plasmonic waveguide modulators. *Nat. Commun*. **6** (2015), doi:10.1038/ncomms9846.

27. M. P. Nielsen *et al*., Adiabatic Nanofocusing in Hybrid Gap Plasmon Waveguides on the Silicon-on-Insulator Platform. *Nano Lett*. **16**, 1410–1414 (2016).

28. V. J. Sorger, R. F. Oulton, J. Yao, G. Bartal, X. Zhang, Plasmonic Fabry-Pérot Nanocavity. *Nano Lett*. **9**, 3489–3493 (2009).

29. R. F. Oulton, G. Bartal, D. F. P. Pile, X. Zhang, Confinement and propagation characteristics of subwavelength plasmonic modes. *New J. Phys*. **10**, 105018 (2008).

30. G. Giovannetti *et al*., Doping graphene with metal contacts. *Phys. Rev. Lett*. **101** (2008), doi:10.1103/PhysRevLett.101.026803.

31. A. Das *et al*., Monitoring dopants by Raman scattering in an electrochemically top-gated graphene transistor. *Nat. Nanotechnol*. **3**, 210 (2008).

32. A. C. Ferrari, D. M. Basko, Raman spectroscopy as a versatile tool for studying the properties of graphene. **8**, 235 (2013).

33. A. C. Ferrari, Raman spectroscopy of graphene and graphite: Disorder, electron–phonon coupling, doping and nonadiabatic effects. *Solid State Commun*. **143**, 47–57 (2007).

34. S. Heeg *et al*., Polarized Plasmonic Enhancement by Au Nanostructures Probed through Raman Scattering of Suspended Graphene. *Nano Lett*. **13**, 301–308 (2013).

35. T. J. Echtermeyer *et al*., Photothermoelectric and Photoelectric Contributions to Light Detection in Metal–Graphene–Metal Photodetectors. *Nano Lett*. **14**, 3733–3742 (2014).

36. J. P. Balthasar Mueller, N. A. Rubin, R. C. Devlin, B. Groever, F. Capasso, Metasurface Polarization Optics: Independent Phase Control of Arbitrary Orthogonal States of Polarization. *Phys. Rev. Lett*. **118** (2017), doi:10.1103/PhysRevLett.118.113901.

37. T. Mueller, F. Xia, M. Freitag, J. Tsang, P. Avouris, Role of contacts in graphene transistors: A scanning photocurrent study. *Phys. Rev. B*. **79** (2009), doi:10.1103/PhysRevB.79.245430.

38. D. Van Tuan *et al.*, Scaling Properties of Charge Transport in Polycrystalline Graphene. *Nano Lett.* **13**, 1730–1735 (2013).

39. B. Huard *et al.*, Transport Measurements Across a Tunable Potential Barrier in Graphene. *Phys. Rev. Lett.* **98** (2007), doi:10.1103/PhysRevLett.98.236803.

40. L. M. Zhang, M. M. Fogler, Nonlinear Screening and Ballistic Transport in a Graphene p – n Junction. *Phys. Rev. Lett.* **100** (2008), doi:10.1103/PhysRevLett.100.116804.

41. M. Freitag, T. Low, F. Xia, P. Avouris, Photoconductivity of biased graphene. *Nat. Photonics.* **7**, 53–59 (2013).

42. L. Lafone, T. P. H. Sidiropoulos, R. F. Oulton, Silicon-based metal-loaded plasmonic waveguides for low-loss nanofocusing. *Opt. Lett.* **39**, 4356 (2014).

43. A. Pospischil *et al.*, CMOS-compatible graphene photodetector covering all optical communication bands. *Nat. Photonics.* **7**, 892–896 (2013).